\documentclass[aps,prl,twocolumn,preprintnumbers,amsmath,amssymb,superscriptaddress]{revtex4-1}
\usepackage{graphicx}
\usepackage{subfigure}
\usepackage{epsfig}
\usepackage{dcolumn}
\usepackage{bm}
\usepackage{ulem}
\usepackage{color}

\def\avg#1{\left\langle#1\right\rangle}
\def\bra#1{\left\langle#1\right|}
\def\ket#1{\left|#1\right\rangle}

\def\be{\begin{equation}}       \def\ee{\end{equation}}
\def\bea{\begin{eqnarray}}      \def\eea{\end{eqnarray}}
\def\ba{\begin{array} }
\def\ea{\end{array} }
\def\bnum{\begin{enumerate} }
\def\enum{\end{enumerate}}

\def\nn{\nonumber}

\def\=>{\Rightarrow}
\def\>{\rightarrow}

\def\eye2{Fathbb{I}}

\def\Eq#1{Eq.~(\ref{#1})}
\def\Fig#1{Fig.~\ref{#1}}

\def\Tr{\mathrm{Tr}}

\renewcommand{\>}{\rangle}

\begin{document}

\title{Solving fermion sign problem in quantum Monte Carlo by Majorana representation}
\author{Zi-Xiang Li}
\affiliation{Institute for Advanced Study, Tsinghua University, Beijing 100084, China}
\author{Yi-Fan Jiang}
\affiliation{Institute for Advanced Study, Tsinghua University, Beijing 100084, China}
\affiliation{Department of Physics, Stanford University, Stanford, CA 94305 China}
\author{Hong Yao}
\email{yaohong@tsinghua.edu.cn}
\affiliation{Institute for Advanced Study, Tsinghua University, Beijing 100084, China}

\begin{abstract}
In this paper, we discover a new quantum Monte Carlo (QMC) method to solve the fermion sign problem in interacting fermion models by employing Majorana representation of complex fermions. We call it ``Majorana QMC'' (MQMC). MQMC simulations can be performed efficiently both at finite and zero temperatures. Especially, MQMC is fermion sign free in simulating a class of
{\it spinless} fermion models on bipartite lattices at half filling and with arbitrary range of
(unfrustrated) interactions.  Moreover, we find a class of $SU(N)$ fermionic models with
{\it odd} $N$, which are sign-free in MQMC but whose sign problem cannot be in solved in other QMC methods such as continuous-time QMC. To the best of our knowledge, MQMC is the {\it first} auxiliary field QMC method to solve fermion sign problem in spinless (more generally, odd number of species) fermion models. We conjecture that MQMC could be applied to solve fermion sign problem in more generic fermionic models.
\end{abstract}
\date{\today}

\maketitle
{\bf Introduction:} Interacting fermionic quantum systems with strong correlations and/or topological properties
have attracted increasing attentions\cite{XGWenbook,Fradkinbook}.
Nonetheless, in two and higher spatial dimensions, strongly interacting quantum systems are generically
beyond the reach of analytical methods in the sense of solving those quantum models
in an unbiased way. As an intrinsically-unbiased numerical method, quantum Monte Carlo simulation plays
a key role in understanding physics of strongly correlated many-body
systems\cite{Scalapino-81, Blankenbecler-81, Rebbi-81, Hirsch-81, Hirsch-85}.
Unfortunately, in simulating fermionic many-body systems, QMC often encounters the notorious
fermion minus-sign problem\cite{Loh-90,Troyer-05}, which arises as a consequence of Fermi
statistics\cite{footnoteFermi}.
Undoubtedly, generic solutions of fermion sign problems would lead to a great leap
forward in understanding correlated electronic systems\cite{Troyer-05}.

Many QMC algorithms are based on converting an interacting fermion
model into a problem of free fermions interacting with background auxiliary
classical fields; the Boltzmann weight is the determinant of free fermion matrix
which is a function of auxiliary fields
and which can be positive, negative, or even complex. In such determinant QMC (DQMC),
when the determinants are rendered to be positive definite, we say a solution to the fermion sign problem
is found. For spinful electrons, conventional strategy of solving fermion sign problem is to find
a symmetric treatment of both spin components of electrons such that the Boltzmann
weight can be written as the product of two real determinants with the same sign and is
then positive definite\cite{Hirsch-86,Hands-00,Wu-05,Berg-12,Assaad-12,Wu-12}.
For spinless or spin-polarized fermion models, it is usually much more difficult to
solve fermion sign problem because the Boltzmann weight contains only a single determinant
and the usual strategy used for even species of fermions cannot be directly applied here.

In this paper, based on Majorana representation of fermions, we propose a genuinely new auxiliary field QMC approach to solve fermion sign problem in spinless fermion models. We observe that each complex fermion can be represented as two Majorana fermions. Consequently, we can express spinless fermion Hamiltonians in Majorana representation and then perform Hubbard-Stratonovich (HS) transformations to decouple interactions by introducing background auxiliary fields. Under certain conditions such as particle-hole symmetry, we can find a symmetric treatment of two species of Majorana fermions, namely the free Majorana fermion Hamiltonian obtained after HS transformations is a sum of two symmetric parts each involving only one species of Majorana fermions, such that the Boltzmann weight is a product of two identical real quantities and is then positive definite.
This is the basic idea of the Majorana approach to solve fermion sign problem in spinless or spin-polarized fermion models which we call ``Majorana QMC'' (MQMC). Note that the MQMC approach proposed here is qualitatively different from the meron-cluster method\cite{Wiese-95, Wiese-99} and fermion bags method\cite{Shailesh-13, Shailesh-14} developed previously, all of which are based on continuous-time QMC (CTQMC)\cite{RMP-11,Shailesh-14,Lei-14,Leiwang-14}. As far as we know, MQMC is the first QMC approach based on auxiliary fields to solve fermion sign problem in a
class of spinless (more generally, odd number of species) fermion models. Moreover, MQMC has an
important advantage: it is much more efficient than continuous-time QMC in simulating models at
low and zero temperatures; the computation-time cost in MQMC scales as $\beta\equiv 1/T$ while
it scales as $\beta^3$ in continuous-time QMC\cite{Shailesh-14} (also see more recent development\cite{troyer-14}).

As an application of the sign-free MQMC algorithm, we have used it to study the charge density wave (CDW) quantum phase transition of the spinless fermion model with repulsive density interactions on the honeycomb lattice with much larger system size ($2L^2$ sites with $L$ up to 24) than previous studies and obtained quantum critical exponents which are in reasonable agreement with renormalization group calculations\cite{unpublished}. We also show that MQMC
can solve the fermion-sign problem in a class of $SU(N=odd)$ models which are beyond
the capability of other QMC methods such as the continuous-time QMC.

\begin{figure}[t]						
\includegraphics[width=6.0cm]{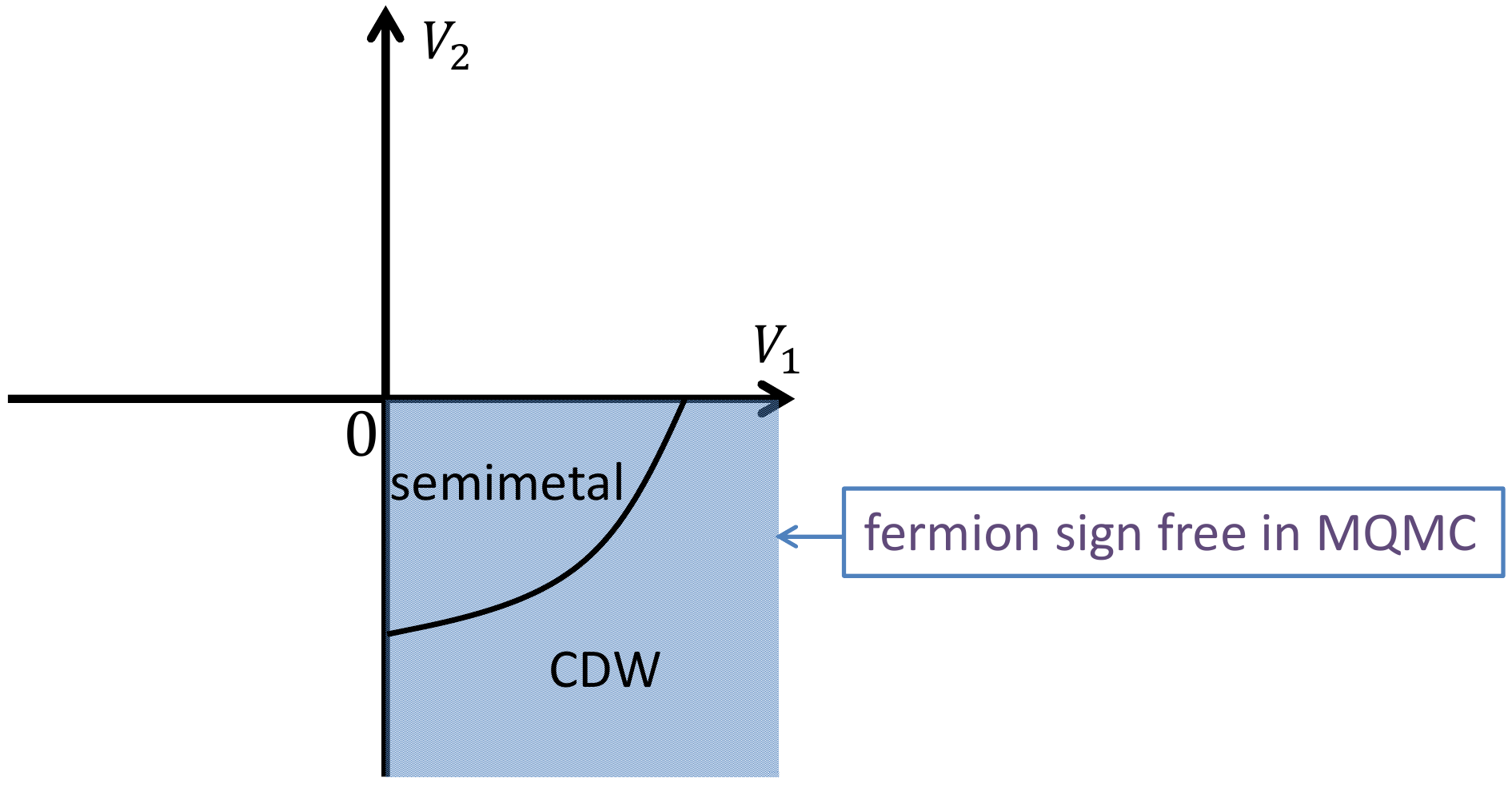}
\caption{The schematic quantum phase diagram of the $t$-$V_1$-$V_2$ spinless fermion model on the honeycomb lattice in the region of $V_1>0$ and $V_2<0$. In this region, MQMC simulations at zero and finite temperatures can be performed efficiently without fermion sign problem. }
\label{pd-dirac}
\end{figure}

{\bf Majorana quantum Monte Carlo:} To explicitly illustrate how MQMC could solve the fermion sign problem in a class of spinless fermion models, we consider the following general Hamiltonian of spinless fermions:
\bea
H &=& H_{0} + H_{\textrm{int}}, \label{H}\\
H_{0} &=& - \sum_{ij}\left[t_{ij}c^{\dagger}_{i}c_{j} + h.c.\right],  \\
H_{\textrm{int}} &=& \sum_{ij} V_{ij}(n_{i}-1/2)(n_{j}-1/2),
\eea
where $c_i^\dag$ creates a fermion on site $i$, $t_{ij}$ represents hopping integral and $V_{ij}$ labels density interaction. As we shall show below, the MQMC is fermion-sign-free when the Hamiltonian in \Eq{H} satisfies the following two conditions: (1) $t_{ij}\neq 0$ {\it only} when $i,j$ belong to different sublattices; (2) $V_{ij}>0$ when $i,j$ belong to different sublattices and $V_{ij}<0$ when $i,j$ belong to same sublattices. With the first condition, it is clear that the model is invariant under particle-hole transformations: $c_i \to (-1)^i c^\dag_i$, where $(-1)^i$ has opposite signs for different sublattices and then describes fermions at half-filling. The lattice in question can be any bipartite lattice such as honeycomb and square lattices in 2D as well as cubic and diamond lattices in 3D. For simplify, we hereafter consider the model with only nearest-neighbor (NN) hopping $t$, NN repulsive interaction $V_{1}$, and next-nearest-neighbor (NNN) attractive interactions $V_{2}$ which we call the $t$-$V_1$-$V_2$ model on the honeycomb lattice (generalizing the MQMC method to models with longer-range hopping/interactions will be straightforward). As shown in \Fig{pd-dirac}, MQMC is fermion sign free in the region where the quantum phase transition between Dirac semimetal and charge density wave (CDW) phases occurs\cite{Lei-14}. (It is interesting to note that the $t$-$V_1$-$V_2$ spinless fermion model on the honeycomb lattice feature very interesting phases including quantum anomalous Hall (QAH) phases\cite{Raghu-08} and pair density wave (PDW) phases\cite{Jian-14}.) 

In statistical physics, a key quantity is the partition function. QMC methods are designed to simulate partition functions in a statistical fashion. For the $t$-$V_1$-$V_2$ model, the partition function after Trotter decomposition is given by
\bea
Z = \Tr\left[e^{-\beta H}\right]\simeq\Tr\left[\prod_{n=1}^{N_\tau} e^{-H_{0}(n)\Delta\tau}e^{-H_{\textrm{int}}(n)\Delta\tau}\right],
\eea
where $n=1,\cdots,N_\tau$ labels the discrete imaginary time, $\Delta\tau N_\tau= \beta$, and the approximation is good for small $\Delta\tau$ or large $N_\tau$. HS transformations can be applied to decouple fermion interactions into non-interacting terms interacting with background auxiliary fields. Usual HS decoupling in density channels normally result in minus sign problem in QMC because the Boltzmann weight is a single determinant. However, we observe that the Hamiltonian can be rewritten in terms of Majorana fermions and there are two species of Majorana fermions. In Majorana representation, complex fermions operators are given by:
\bea
c_{i}=\frac{1}{2}(\gamma_{i}^{1}+i\gamma_{i}^{2}), ~~c^{\dagger}_{i}=\frac{1}{2}(\gamma_{i}^{1}-i\gamma_{i}^{2}),
\eea
which enable us to rewrite the Hamiltonian as follows:
\bea
H_0&=& \sum_{\avg{ij}} \frac{it}{2}(\gamma^{1}_{i}\gamma^{1}_{j} + \gamma^{2}_{i}\gamma^{2}_{j}), \nn\\
H_{\textrm{int}} &=& - \frac{V_{1}}{4}\sum_{\avg{ij}}(i \gamma_{i}^{1}\gamma_{j}^{1})(i \gamma_{i}^{2}\gamma_{j}^{2}) - \frac{V_{2}}{4}\sum_{\avg{\avg{ij}}} ( i \gamma_{i}^{1}\gamma_{j}^{1})(i \gamma_{i}^{2}\gamma_{j}^{2}), \nn
\eea
where gauge transformations $c_{i}\rightarrow i c_{i}$ for $i$ in only one sublattice were implicitly made so that $H_0$ can be written symmetrically in the two components of Majorana fermions. Now, it is clear that we should perform HS transformations in Majorana hopping channels instead of density channels as done in usual QMC methods. Explicitly, HS transformations for interactions in $H_{int}$ in MQMC are given by
\bea
&& e^{\frac{V_{1}\Delta\tau}{4}(i \gamma_{i}^{1}\gamma_{j}^{1})(i\gamma_{i}^{2}\gamma_{j}^{2})}
= \frac{1}{2}\sum_{\sigma_{ij}=\pm 1}e^{\frac12\lambda_{1} \sigma_{ij} (i \gamma_{i}^{1} \gamma_{j}^{1} + i \gamma_{i}^{2}\gamma_{j}^{2})-\frac{V_1\Delta\tau}{4}}, \label{V1sign}~~~\\
&&e^{\frac{V_{2}\Delta \tau}{4}(i \gamma_{i}^{1}\gamma_{j}^{1})(i\gamma_{i}^{2}\gamma_{j}^{2})} = \frac{1}{2}\sum_{\sigma_{ij}=\pm 1}e^{\frac12\lambda_{2} \sigma_{ij}  (i \gamma_{i}^{1} \gamma_{j}^{1}-i \gamma_{i}^{2}\gamma_{j}^{2})+\frac{V_2\Delta\tau}{4}},~~~\label{V2sign}
\eea
where $\lambda_1$ and $\lambda_2$ are constants determined through $\cosh \lambda_{1} = e^{\frac{V_{1}\Delta\tau}{2}}$ and $\cosh \lambda_{2} = e^{\frac{-V_{2}\Delta\tau}{2}}$, respectively. Note that in \Eq{V2sign} the signs of $\gamma^{1}$ hopping terms are opposite to $\gamma^{2}$ hopping terms in the HS decompositions of NNN interaction because $V_2<0$. The same signs are obtained for decoupling of NN interactions in \Eq{V1sign} because $V_1>0$. It is now clear that the free fermion Hamiltonian after the HS transformations is a sum of two parts each of which involves only one component of Majorana fermions. This makes MQMC simulations sign-problem free because the Boltzmann weight can be positive definite, which we shall show below.

Note that auxiliary fields $\sigma_{ij}(n)$ should be introduced independently for each discrete imaginary time $n$. As a result, the partition function is a sum over Boltzmann weight which is a function of auxiliary field configurations in space-time, as given by
\bea
Z=\sum_{\{\sigma\}} W(\{\sigma\}).
\eea
Up to an unimportant constant the Boltzmann weight $W(\{\sigma\})$ is given by
\bea\label{trace0}
W(\{\sigma\})= \Tr\left[ \prod_{n=1}^{N_\tau} e^{\sum_{a=1}^2 \frac14\widetilde \gamma^a h^a (n) \gamma^a} \right],
\eea
where $\widetilde \gamma^a$ represents the transpose of $\gamma^a$ and $h^a(n)$ is a $N\times N$ matrix ($N$=the number of lattice sites) is given by
\bea
h^a_{ij}(n) \!=\! i\left[t\Delta\tau \delta_{\avg{ij}}+\lambda_1\sigma_{ij}(n)\delta_{\avg{ij}} \pm \lambda_2\sigma_{ij}(n)\delta_{\avg{\avg{ij}}} \right],~~~~~
\eea
where $\delta_{\avg{ij}}=\pm 1$ if $ij$ are NN sites and 0 otherwise; similarly $\delta_{\avg{\avg{ij}}}=\pm 1$ only if $ij$ are NNN sites. Now, we can trace out the Majorana fermions since they are free, as shown in Supplemental Material. Because the two components of Majorana fermions are decoupled, tracing out Majorana fermions can be done independently and the Boltzmann weight is a product of two factors:
\bea\label{trace1}
W(\{\sigma\})= W_1(\{\sigma \}) W_2(\{\sigma\}),\nonumber
\eea
where
\bea\label{trace2}
W_a(\{\sigma\})= \left\{\det\bigg[ \mathbb{I}+\prod_{n=1}^{N_\tau} e^{h^a(n)}
 \bigg]\right\}^{\frac{1}{2}}.~~
\eea
Note that there is sign ambiguity when taking a square root above, similar to the case of Pfaffian as a square root of determinants.

{\bf Fermion sign free:} Now we prove that the Boltzmann weight is positive definite by showing that $W_1(\{\sigma\})=W^\ast_2(\{\sigma\})$. A key observation is that the Hamiltonian $\hat h^1(n)\equiv \widetilde \gamma^1 h^1(n)\gamma^1$ of Majorana fermions $\gamma^1$ can be mapped to a Hamiltonian identical to $\hat h^2(n)\equiv \widetilde \gamma^2 h^2(n)\gamma^2$ by the following time-reversal transformation $\Theta=T K$, where $K$ is the complex conjugation and $T$ is given as below:
\bea\label{gt}
T&:& \gamma_i^1 \to (-1)^i \gamma_i^1.
\eea
Namely, $\widetilde \gamma^1 h^1 (n)\gamma^1 \to \widetilde \gamma^2 h^2(n) \gamma^2$ under the time reversal transformation $\Theta$. Because the time-reversal transformation complex conjugates the results of tracing out Majorana fermions, we obtain
\bea
W_{1}(\{\sigma\})= W_2^{*}(\{\sigma\}),
\eea
which renders the Boltzmann weight $W(\{\sigma\}) = W_{1}(\{\sigma\}) W_2(\{\sigma\})=W_{1}(\{\sigma\}) W^\ast_1(\{\sigma\})\ge 0$ for any auxiliary field  configuration $\{\sigma\}$. Explicitly, it is
\bea\label{boltzmann}
W(\{\sigma\})=\left|\det\Big[ \mathbb{I}+\prod_{n=1}^{N_\tau} e^{h^a(n)}
\Big]\right|,
\eea
where $a=1$ or $2$, which gives rise to the same result. This proves that the MQMC algorithm can solve fermion sign problem in such class of models consisting of spinless fermions. It is the central result in this paper.

{\bf Projector MQMC:} The MQMC algorithm above simulates finite-temperature partition function in the grand canonical ensemble by computing the trace in \Eq{trace0}. If one is interested in ground state properties, it is of advantage to use the projector algorithm to carry out QMC \cite{Sugiyama-86,Sorella-89,White-89} since projector QMC is often more efficient than finite-temperature QMC. The expectation value of an operator $O$ in the ground state is given by
\begin{equation}
\frac{\bra{\psi_{0}} O \ket{\psi_{0}}}{\avg{\psi_{0}\mid \psi_{0}}} = \lim_{\theta \rightarrow \infty } \frac{\bra{\psi_{T}} e^{-\theta H} O e^{-\theta H} \ket{ \psi_{T}}}{\bra{\psi_{T}} e^{-2 \theta H}  \ket{\psi_{T}}},
\end{equation}
where $\ket{\psi_0}$ is the ground state and $\ket{\psi_T}$ is a trial wave function which we assume has a finite overlap with the true ground state. Here, $Z_T\equiv \bra{\psi_{T}} e^{-2 \theta H}  \ket{\psi_{T}}$ plays the role of usual partition functions and need to be expressed as a sum of Boltzmann weights. In practice, a Slater-determinant wave function describing non-interacting fermions is often chosen as the trial wave function in projector QMC:
\begin{equation}
 \ket{\psi_{T}}  = \prod_{\alpha=1}^{N_{f}} (c^{\dagger} P)_\alpha  \ket{0},\label{projector}
\end{equation}
where $P$ is a $N\times N_f$ matrix ($N_{f}$ labels the number of fermions in question).  Usually, $\ket{\psi_T}$ is an eigenvector of the non-interacting part of the Hamiltonian in question, namely $H_0$ in \Eq{H}. In Majarana representation of fermions, $\gamma^1$ and $\gamma^2$ Majorana fermions are decoupled in $H_0$; consequently $\ket{\psi_T}=\ket{\psi^1_T}\otimes \ket{\psi^2_T}$. By introducing similar HS transformations and auxiliary fields $\{\sigma\}$ as above, the ``partition function'' is obtained a sum of Boltzmann weight $W(\{\sigma\})$ over auxiliary field configurations: $Z_T=\sum_{\{\sigma\}} W(\{\sigma\})$. Since $\gamma^1$ and $\gamma^2$ Majorana fermions are decoupled after the HS transformation, we again obtain $W(\{\sigma\})=W_1(\{\sigma\}) W_2(\{\sigma\})$, where
\bea
W_a(\{\sigma\})=\bra{\psi^a_T}\left[ \prod_{n=1}^{N_\tau} e^{ \frac14\widetilde \gamma^a h^a (n) \gamma^a} \right]\ket{\psi^a_T}.
\eea
Similarly, $W_1(\{\sigma\})=W^\ast_2(\{\sigma\})$ because of the time reversal symmetry $\Theta$. As shown in the Supplemental Material, the Boltzmann weight is given by
\bea\label{projector}
W(\{\sigma\})=\left|\det\bigg\{ P_a^\dag \bigg[ \prod_{n=1}^{N_\tau} e^{h^a(n)} \bigg]P_a \bigg\}\right|,~~~~
\eea
where $a=1$ or 2 and $P_a$ is the projection matrix constructed from $\ket{\psi^a_T}$. Consequently, the projector MQMC is also free from fermion sign problem for a class of spinless fermion models.

{\bf Physical observables in MQMC:} One important advantage of auxiliary-field QMC algorithms is that physical observables can be obtained conveniently. For instance, time and space dependent Green's function can be computed directly in DQMC algorithm. We show below that both at finite and zero temperature the computation of physical observables in MQMC is similarly convenient as that in DQMC algorithm.

In QMC, physical observables can be related to single-particle Green's function: $G_{ij} = \langle  c^\dag_ic_j\rangle $, where the average is done stochastically over auxiliary field configurations. In Majorana representation, it is given by
\bea\label{cor}
\big\langle c^\dag_i c_j\big\rangle  = \frac14\Big[\avg{ \gamma^{1}_i \gamma^{1}_j} +\avg{ \gamma^2_i\gamma^2_j }\Big],
\eea
where we used the results of $\avg{\gamma^1_i \gamma^2_j}=0$ which is a consequence of the decoupling of the two species of Majorana fermions after the HS transformation. To obtain the Green's functions, we only need to compute $\avg{\gamma^1_i \gamma^1_j}$ and $\avg{\gamma^2_i\gamma^2_j}$. Because the two species of Majorana fermions are related by the time reversal symmetry $\Theta$, we obtain $W_{1}(\{\sigma\}) = W^\ast_{2}(\{\sigma\})$.
It is straightforward to evaluate the equal-time Majorana Green's function $\avg{\gamma^a_i \gamma^a_j}$ in finite temperature MQMC:
\bea
G^a_{ij}&=&\sum_{\{\sigma\}} W(\{\sigma\}) \avg{\gamma^a_i \gamma^a_j}_\sigma,  \nn\\
&=& \frac12 \sum_{\{\sigma\}} W(\{\sigma\}) \Big[\mathbb{I}+\prod_{n=N_\tau}^{1} e^{-h^a(n)}\Big]^{-1}_{ji},~~~~~
\eea
where the factor 1/2 above comes from the nature of Majorana fermions. Employing Wick's theorem for each configuration $\{\sigma\}$, higher order correlation functions, including density-density and pair-pair correlations, can be obtained
from single-particle Green's functions. For instance, the equal-time density-density correlations are given by
$\big\langle(c_{i}^{\dagger}c_{i}-\frac12)(c^{\dagger}_{j}c_{j}-\frac12)\big\rangle_\sigma
= \frac14\avg{\gamma^{1}_{i}\gamma^{1}_{j}}_\sigma  \avg{\gamma^{2}_{i}\gamma^{2}_{j}}_\sigma$.

It is increasingly realized that quantum entanglement could play a key role in understanding quantum many-body systems\cite{Cardy-04,Kiteav-06,Levin-06,Haldane-08,Eisert-10}.
Quantum entanglement is partially characterized by entanglement entropy,
including the von Neumann entropy $S_{vN}=-\Tr[\rho_A\log\rho_A]$ and Renyi entropy $S_{n} = -\frac{1}{n-1}\log[\Tr(\rho_{A}^{n})]$ where $\rho_A$
is the reduced density matrix of subregion $A$.
Even though it is still challenging for auxiliary-field QMC algorithms to evaluate von Neumann entropy, it was shown recently that DQMC can provide an efficient way to evaluate Renyi entropy
by simulating the reduced density matrix $\rho_A$ expressed in terms of Green's function\cite{Grover-13,Assaad-14}. Because MQMC is able to compute Green's functions efficiently, Renyi entropy can be calculated accurately in MQMC algorithm as long as it is fermion sign free.

\begin{figure}[t]						
\centering
\subfigure[]{\includegraphics[width=4.36cm]{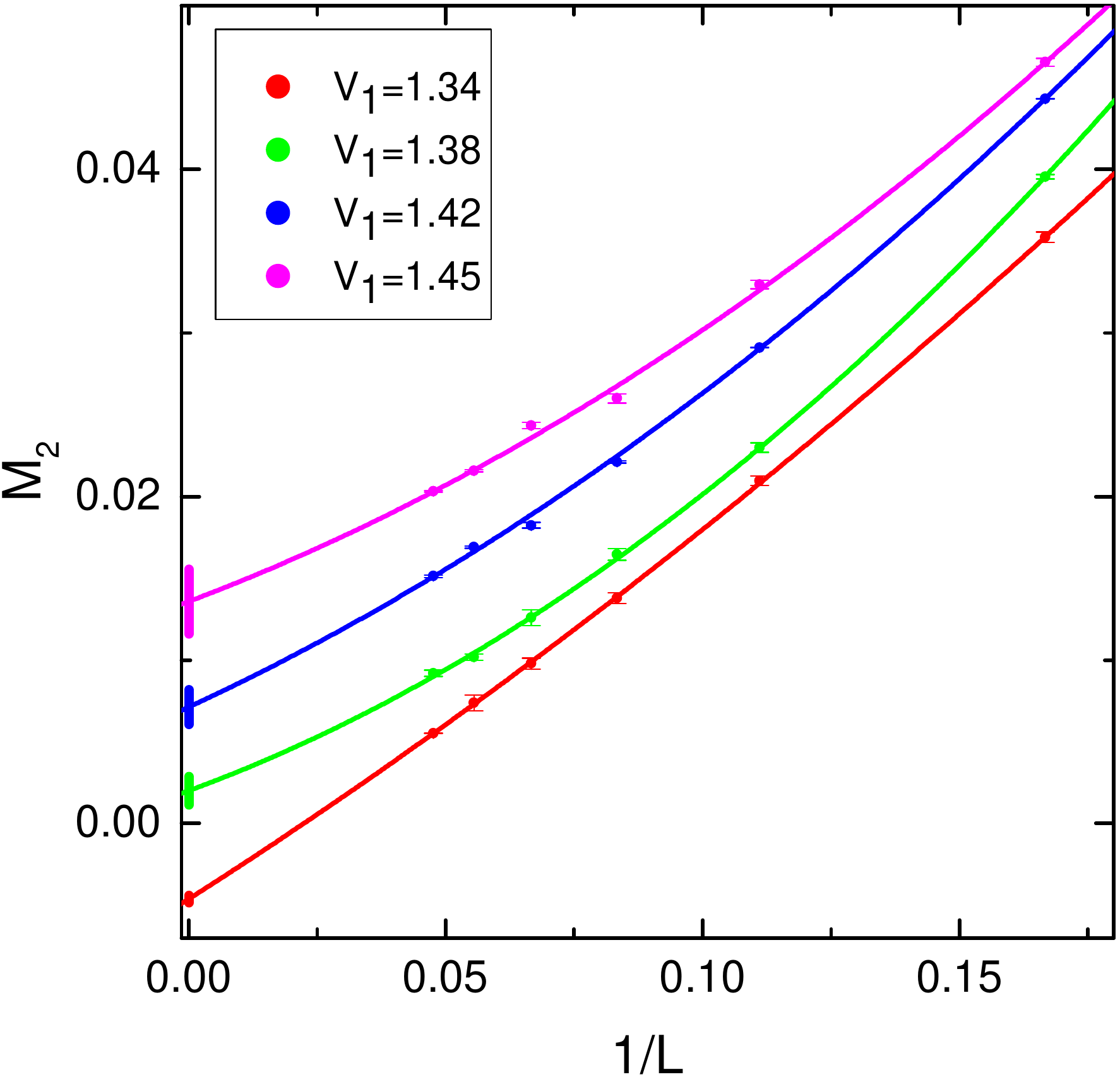}}
\subfigure[]{\includegraphics[width=4.19cm]{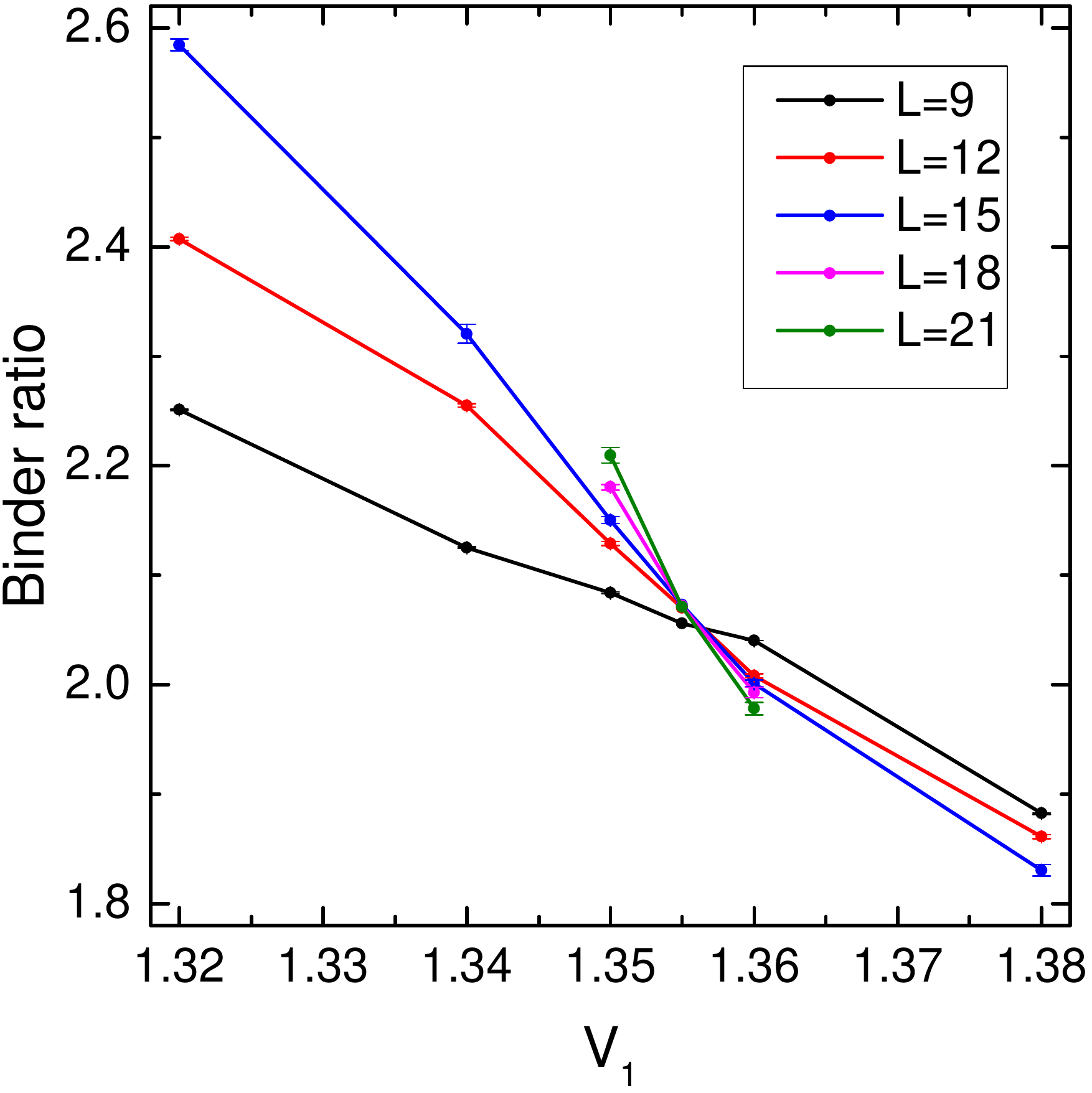}}
\caption{(a) Finite-size scaling of the CDW structure factor $M_2$ obtained in the projector (zero-temperature) MQMC simulations for various $V_1$ and $L=9,12,15,18,21$. It is clear that the phase transition between the semimetal and the CDW phase occurs when $V_1$ is between $1.34$ and $1.38$. The error bars for measured quantities are show explicitly and they are negligibly small. (b) The Binder ratios $B\equiv M^4/M_2^2$ for various $V_1$, including $V_1=1.355$, and various $L=9\sim 21$, are plotted. From crossing of Binder ratios, we conclude that the critical value of $V_1$ for the CDW transition is $V_{1c}=1.355\pm 0.001$. }
\label{scaling}
\end{figure}

{\bf Numerical Results:} We performed highly-accurate projector MQMC simulations to study the $t$-$V_1$-$V_2$ model on the honeycomb lattice at zero temperature. For simplicity, we set $t=1$, $V_2=0$, and then vary $V_1$ to find the critical value of $V_1$, above which the system develops a finite CDW ordering at zero temperature. To measure the CDW order parameter $\Delta_\textrm{CDW}$, we calculate CDW structure factor at finite lattice size:
\bea
M_2=\sum_{ij}\frac{\eta_i\eta_j}{N^2} \big\langle(n_i-\frac12)(n_j-\frac12)\big\rangle,
\eea
where $\eta_i=+1 (-1)$ on $A (B)$ sublattice and $N=2\times L\times L$ is the total number of sites.
It is obvious that $\lim_{L\to\infty} M_2 = \Delta_\textrm{CDW}^2$. The simulations are done for
lattices up to $L=21$ which is substantially larger than the one in Ref. \cite{Lei-14} indicating
that our MQMC algorithm is quite efficient. As shown in \Fig{scaling}(a), we obtain $\Delta_\textrm{CDW}^2$
through finite-size scaling of the measured $M_2$ on lattices of $L=9,12,15,18,21$. For instance,
$\Delta_\textrm{CDW}\approx 0.17\pm 0.01$ at $V_1=1.42$. It is clear that the critical value of
$V_1$ separating the semimetal and CDW phases is between 1.34 and 1.38. To obtain the critical
value of $V_1$ more accurately, we calculate the Binder ratio defined as $B=\frac{M_4}{M_2^2}$ for
various $V_1$ and $L$, where $M_4=\sum_{ijkl}\frac{\eta_i\eta_j\eta_k\eta_l}{N^4}
\big\langle(n_i-\frac12)(n_j-\frac12)(n_k-\frac12)(n_l-\frac12)\big\rangle$.
At the putative critical point, the Binder ratios for different $L$ should cross.
As shown in \Fig{scaling}(b), the Binder ratios for $L=12,15,18,21$ indeed cross nearly the
same point when $V_1=1.355$. Consequently, we conclude that the critical value $V_{1c}=1.355\pm 0.001$.

The critical exponents and universality class at the phase transition\cite{sachdevbook,Herbut-14}
have been analyzed through even larger-scale MQMC simulations by us\cite{unpublished} .
Because the CPU time-cost scales linearly with $\beta$, we were able to perform the MQMC simulations on much larger
system size ($L_{max} =24$)\cite{unpublished} than the one studied by CTQMC ($L_{max} =15$ there)\cite{Lei-14};
consequently the critical exponents obtained by MQMC are reasonably consistent with RG calculations.

{\bf Other models sign-free in MQMC:} We have shown that MQMC, as a new auxiliary field QMC approach, can solve fermion sign problem in a class of {\it spinless} fermion models by utilizing Majorana representation of complex fermions. It will be straightforward to generalize the current MQMC algorithm to solve the fermion sign problem in interacting fermion models with more than one fermion species. Such MQMC fermion-sign free models include the $SU(N=
odd)$ negative-$U$ Hubbard model on bipartite lattices whose Hamiltonian is 
\bea
H=-t\sum_{\avg{ij}} \left[\sum_{\alpha=1}^Nc^\dag_{i\alpha}c_{j\alpha}\!+\!h.c.\right]
+U\sum_i\Big[n_i-\frac{N}2\Big]^2, ~~
\eea
where $U<0$ and $n_i=\sum_\alpha c^\dag_{i\alpha}c_{i\alpha}$. This model on the honeycomb lattice has a similar
semimetal to CDW transition even though the quantum critical exponents can depend on $N$.

More importantly, we can show that the following $SU(N=odd)$ fermionic model
\bea
H\!=\!-t\sum_{\avg{ij}} \left[\sum_{\alpha=1}^Nc^\dag_{i\alpha}c_{j\alpha}\!+\!h.c.\right]
\!- \!J\sum_{\avg{ij}}\left[c^\dag_{i\alpha}c_{j\alpha}\!+\!h.c.\right]^2,~~~~~
\eea
is sign-free in MQMC when the lattice is bipartite and $J>0$. It is worth to stress that this
class of $SU(N)$ models are sign-free {\it only} in the MQMC method but encounter sign-problem in
other QMC methods such as CTQMC\cite{Shailesh-14, Lei-14}. This shows that the MQMC algorithm discovered by us can solve the fermion-sign of models which go beyond those solvable by CTQMC and other conventional QMC methods.

{\it Acknowledgement:} We thank Fakher Assaad, Alexei Kitaev, Ziyang Meng, and Lei Wang for helpful discussions. This work is supported in part by the National Thousand-Young-Talents Program (HY) and the NSFC under grant No. 11474175 (ZXL, YFJ, and HY).

\begin{widetext}

\section{SUPPLEMENTAL MATERIALS}
\renewcommand{\theequation}{S\arabic{equation}}
\setcounter{equation}{0}
\subsection{Appendix A: Trace involving Majorana fermions}
Now we show that the trace of exponentials of bilinear Majorana fermion operators can be expressed as the square root of a determinant (nonetheless, it is formally not a Pfaffian as shown below).

First, we evaluate the trace of a single exponential of bilinear Majorana fermion operators. Suppose $\hat h = \frac14\sum_{ij} \gamma_{i}h_{ij}\gamma_{j}$ where $h_{ij}=-h_{ji}$ and we need to compute $\Tr[e^{-\Delta\tau \hat h}]$. After diagonalizing $\hat h=\sum_{a=1}^{N/2} [\frac12\epsilon_{a}c^{\dagger}_{a}c_{a} - \frac12\epsilon_{a}c_{a}c^{\dagger}_{a}]$, where $\pm \epsilon_a$ are eigenvalues of the $N\times N$ matrix $h$, it is clear that the trace is given by
\bea
\Tr [e^{-\Delta\tau \frac14\widetilde \gamma h \gamma}] = \prod_{a=1}^{N/2}( e^{ \frac12\Delta \tau \epsilon_{a}} + e^{ -\frac12\Delta \tau \epsilon_{a}}),
\eea
which can be reexpressed as the square root of a determinant:
\bea
\Tr [e^{-\Delta \tau \frac14\widetilde \gamma h \gamma}] &=& \Big[\det\big(e^{ \frac12\Delta \tau h}+ e^{- \frac12\Delta \tau h}\big)\Big]^{\frac12},\\
&=&\Big[\det\big(\mathbb{I}+e^{\Delta \tau h}\big)\Big]^{\frac12}
\eea
Intuitively, the square root originates from the fact that Majorana fermions carry only half of degrees of freedom of corresponding Hamiltonian in terms of complex fermions.

Then we show that the product of exponentials of bilinear Majorana fermion operators can be grouped into a single exponential of a bilinear Majorana fermion form. Suppose $U=\prod_n e^{-\Delta\tau\hat h(n)}$, where $\hat h(n)=\frac14 \sum_{ij} \gamma_i h_{ij}(n) \gamma_j$, and we would like to evaluate $\Tr[U]$. By observing that $e^{-\Delta\tau \hat h(n)} \gamma_{i} e^{\Delta\tau \hat h(n)} = \sum_{j} \gamma_{j} [e^{-\Delta \tau h(n)}]_{ji}$, we obtain
\bea
U\gamma_{i}U^{-1} &=& \sum_{j}\gamma_{j}\Big[\prod_{n}e^{-\Delta \tau h(n)}\Big]_{ji},\\
&\equiv &\sum_{j} \gamma_{j} \Big[e^{-\Delta\tau h'}\Big]_{ji},
\eea
where a $N\times N$ matrix $h'$ is defined. Accordingly, we introduce bilinear Majorana fermion operators: $\hat h'=\frac14 \sum_{ij} \gamma_i h'_{ij} \gamma_j$.  Now, we can show that the trace of the product of exponentials of the Majorana fermions bilinear operator is given by the square root of a determinant:
\bea
\Tr\bigg[\prod_{n}e^{-\Delta \tau \frac14\widetilde\gamma h(n) \gamma} \bigg]
&=&\Tr\bigg[e^{-\Delta \tau \frac14 \widetilde\gamma h' \gamma}\bigg],\\
&=&\bigg[\det\Big( \mathbb{I}+e^{-\Delta \tau h'} \Big)\bigg]^{\frac12},\\
&=& \bigg\{\det\Big[\mathbb{I}+ \prod_n e^{-\Delta \tau h(n)}\Big]\bigg\}^{\frac{1}{2}},
\eea
which proves the result in Eq. (11) of the main text.

\subsection{Appendix B: Projector QMC in Majorana representation}
We now prove the result in Eq. (19) of the main text. To compute $W_a(\{\sigma\})$, we compute its square first as follows:
\bea
W^2_a(\{\sigma\})&=&\bra{\psi^a_T}\left[ \prod_{n=1}^{N_\tau} e^{ \frac14\widetilde \gamma^a h^a (n) \gamma^a} \right]\ket{\psi^a_T}\bra{\phi^a_T}\left[ \prod_{n=1}^{N_\tau} e^{ \frac14\widetilde \eta^a h^a (n) \eta^a} \right]\ket{\phi^a_T},\\
&=&\bra{\psi^a_T\otimes \phi^a_T} \left[ \prod_{n=1}^{N_\tau} e^{ \frac14\widetilde \gamma^a h^a (n) \gamma^a + \frac14\widetilde \eta^a h^a (n) \eta^a} \right] \ket{\psi^a_T\otimes \phi^a_T},
\eea
where $\eta^a$ are the ``ghost Majorana fermions'' which are independent from $\gamma^a$ but have the same Hamiltonian and ground state wave function as $\gamma^a$. When combining $\eta^a_i$ and $\gamma^a_i$ into complex fermions $d_j\equiv (\gamma^a_j + i \eta^a_j)/2$, we obtain
\bea
W^2_a(\{\sigma\}) &=& \bra{\psi^a_T\otimes \phi^a_T} \left[ \prod_{n=1}^{N_\tau} e^{ d^\dag h^a (n) d } \right] \ket{\psi^a_T\otimes \phi^a_T},\\
&=& \det\bigg\{P_a^\dag \Big[\prod_{n=1}^{N_\tau} e^{ h^a (n) } \Big] P_a \bigg\},
\eea
where $P_a$ is a $N\times N_f$ projector matrix defined through $\ket{\psi^a_T\otimes \phi^a_T}=\prod_\alpha (d^\dag P_a)_\alpha \ket{0}$. Because $W_1(\{\sigma\}) = W^\ast_2(\{\sigma\}) $, we prove Eq. (19) as follows:
\bea
W(\{\sigma\})=\left|\det\bigg\{ P_a^\dag \bigg[ \prod_{n=1}^{N_\tau} e^{h^a(n)} \bigg]P_a \bigg\}\right|.~~~~
\eea

\subsection{Appendix C: Proof of fermion-sign free in a class of $SU(N=odd)$ fermionic model with bond interactions}
We now prove in details that the fermionic model with $N=odd$ fermion species described by Eq. (24) does not encounter fermion-sign problem in our MQMC algorithm. Both the hopping term and the interaction term in Eq. (24) can be rewritten in terms of Majorana fermions:
\bea
H_0&=& -t \sum_{\avg{ij}} ( c^{\dagger}_{i\alpha}c_{j\alpha} + h.c) = \sum_{\avg{ij}} \frac{it}{2}(\gamma^{1}_{i\alpha}\gamma^{1}_{j\alpha} + \gamma^{2}_{i\alpha}\gamma^{2}_{j\alpha}), \nn\\
H_{int}&=& -J\sum_{\avg{ij}}(c^{\dagger}_{i\alpha}c_{j\alpha} + h.c)^2=  -\frac{J}{4} \sum_{\avg{ij}} ( i \gamma_{i\alpha}^1 \gamma_{j\alpha}^1 + i \gamma_{i\alpha}^1 \gamma_{j\alpha}^2)^2
\eea
where gauge transformations $c_i \rightarrow ic_i$ for $i$ in one sublattice are implicitly made. Then, we can perform a similar Hubbard-Stratonovich (HS) transformation on the bond interactions:
\bea
e^{\frac{J\Delta_\tau}{4} (  \sum_{\alpha=1}^{N} i \gamma^1_{i\alpha}r^2_{j\alpha} + i \gamma^2_{i\alpha}\gamma^2_{j\alpha})^2}= \frac{1}{2}\sum_{\sigma_{ij} = \pm 1} e^{ \lambda \sigma_{ij} ( \sum_{\alpha=1}^{N} i \gamma_{i\alpha}^1 \gamma_{j\alpha}^1 + i \gamma_{i\alpha}^2 \gamma_{j\alpha}^2 )}, ~~~~
\eea
where $\lambda$ is a constant satisfying $ \cosh \lambda = e^{\frac{J \Delta_\tau}{2}}$. For $J>0$, $\lambda$ is a real number. After the HS transformation, it is clear that Hamiltonian is a sum of two parts each of which involves only one component of Majorana fermions ($\gamma^1_{i\alpha}$ and $ \gamma^2_{i\alpha}$):
\bea
h^a_{ij,\alpha}(n) \!=\! i\left[t\Delta\tau \delta_{\avg{ij}}+\lambda\sigma_{ij}(n)\delta_{\avg{ij}} \right],~~~~~
\eea
So Boltzmann weight can be decoupled as the product of $2N$ identical parts:
\bea
W(\{\sigma\}) = \left[W_1(\{\sigma\})\right]^{2N}.
\eea
where $W_1(\{\sigma\})$ is the Boltzmann weight obtained through tracing out one component of the Majorana fermions, say $\gamma^1_{i1}$. Moreover, each part of Hamiltonian is invariant under this anti-unitary time-reversal transformation $\Theta=TK$ where $T$ is given as below:
\bea\label{gt}
T&:& \gamma_{i\alpha}^a \to (-1)^i \gamma_{i\alpha}^a.
\eea
As a result, the Boltzmann weight $ W_1(\{\sigma\})$ is real and consequently $W(\{\sigma\})>0$.  This proves that the fermionic model with $N=odd$ fermion species described by Eq. (24) can be fermion-sign free in MQMC simulations.
\end{widetext}


\begin{thebibliography}{10}
\bibitem{footnoteFermi} A negative sign in quantum magnets with frustrated superexchange interactions is also a consequence of the fermionic minus sign of the superexchange process.
\end{thebibliography}
\end{document}